# Discovery of Proteomics based on Machine Learning

Biao He and Baochang Zhang

*Machine Perception Laboratory,*
*School of Automation Science and Electrical Engineering*
*Beihang University*
Beijing,100191, China

hebiaobuaa@gmail.com; bczhang@buaa.edu.cn

Yan Fu

*Academy of Mathematics and Systems Science*

*China Academy of Sciences*

Beijing, China

yfu@amss.as.cn

*Abstract* – **The ultimate target of proteomics identification is to identify and quantify the protein in the organism. Mass spectrometry (MS) based on label-free protein quantitation has mainly focused on analysis of peptide spectral counts and ion peak heights. Using several observed peptides (proteotypic) can identify the origin protein. However, each peptide's possibility to be detected was severely influenced by the peptide physicochemical properties, which confounded the results of MS accounting. Using about a million peptide identification generated by four different kinds of proteomic platforms, we successfully identified >16,000 proteotypic peptides. We used machine learning classification to derive peptide detection probabilities that are used to predict the number of trypic peptides to be observed, which can serve to estimate the absolutely abundance of protein with highly accuracy. We used the data of peptides (provides by CAS lab) to derive the best model from different kinds of methods. We first employed SVM and Random Forest classifier to identify the proteotypic and unobserved peptides, and then searched the best parameter for better prediction results. Considering the excellent performance of our model, we can calculate the absolutely estimation of protein abundance.**

## I. INTRODUCTION

One of the fundamental target of modern biology is to figure out the relations between the context of the structure and function of genomic information. Different mass spectrometry techniques attempt to provide qualitative results that describe the relations. Isotope labeling and fluorescent labeling techniques have been widely used in quantitative analysis of proteins. However, researcher are turning to label-free methods because it is faster and simpler [1] [2] [3]. Peptides are produced by enzymatic digestion of protein mixture and then put these peptides for training [4].

Label-free method makes use of the number of MS observed tryptic peptides to estimate the relative quantity of a protein [5]. However, spectral counting could be confounded by the possibility of a peptide to be observed [6] [7]. A series of researches discovered that we can identify the protein based on the observance of one or a few preferentially detected peptides [1]. And some researches also discovered that different kinds of peptides' likelihood to be detected can varies from others. Peptide physicochemical properties can affect final MS detection on account of many factors such as length of peptide, mass, average flexibility indices, net charge and other properties which can affect peptide observance [8]. This variability should be taken into consideration of quantity estimation, otherwise errors would happen in assessing the abundance of protein.

To estimate the quantity of protein, we could use the number of detected peptides to indicate the abundance. We used two classifiers to distinguish the peptides to two kinds, called proteotypic and unobserved. We referred these data and features from [1]. And these peptides are produced by four platforms from CAS lab (Chinese Academy of Sciences). In addition, considering the different platforms, we should make the classification separately. We can derive a likelihood for each peptide for its origin protein, then if we are trying to identify and quantify a new protein, the likelihood can be considerably useful to achieve accurate prediction.

## II. FEATURE EXTRACTION

Experiments showed that some peptides are more easily identified than others. Therefore we call the preferentially observed peptides as proteotypic peptide. The existence of proteotypic peptides raised a question that what properties distinguish frequently observed peptides from peptides that are present in the same sample of protein but remain unidentified? And what properties of peptide can be applied into all living organisms. Moreover, if we explored the special properties out, is it possible to predict whether the peptide is proteotypic using the protein's sequence? Towards these goals, we extracted the proteotypic peptides from four different platforms and featured them with physicochemical property.

To identify the properties governing a peptide's proteotypic inclination, we have evaluated 544 different physicochemical property scales for amino acids, including hydrophobicity index, residue volume and transfer free energy to surface [8]. Considering both the sum and the average value could contribute to the propensity at the same time, we employed both value to describe peptides, resulting in 1088-dimensional property vector per peptide (as shown in Fig. 1).

| Peptide Example | K | L | I | G | D | Total | Aver |
|---|---|---|---|---|---|---|---|
| Amino acid composition | 0.68 | 0.98 | 1.02 | 0 | 0.76 | 3.44 | 0.688 |
| Relative mutability | 6.6 | 7.4 | 4.5 | 8.4 | 5.5 | 32.4 | 6.48 |
| Melting point | 56 | 40 | 96 | 49 | 106 | 347 | 69.4 |
| Optical rotation | 224 | 337 | 284 | 290 | 270 | 1405 | 281 |
| Steric parameter | 14.6 | -11 | 12.4 | 0 | 5.05 | 21.05 | 4.21 |

Fig. 1 The genuine and imposter distance distribution

## III. METHODS

*Method #1: Support Vector Machine*

SVM is a computationally effective supervised learning technique which is widely used in pattern recognition and machine learning projects. This method derived from some of knowledge of statistical learning theory regarding controlling the generalization abilities of a learning machine.

A support vector machine establishes a hyper-plane to classify the given pattern. In addition, this approach introduced a transformation method to make better performance, kernel functions: the input feature vector space by applications of a non-linear function can be transformed into a high-dimensional space where the best hyper-plane can be learnt. This can improve simple SVM method to solve much more complex classification between sets which are not convex at all in the original space [9] [10].

The basic idea of SVM classifier can be illustrated by Fig.2.

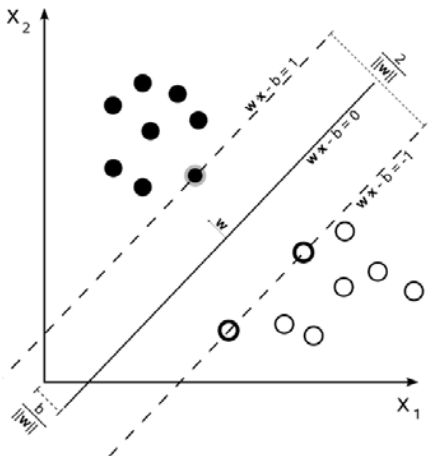

Fig. 2 The hyper-plane and margin for SVM

This figure shows the simplest case that two kinds of vectors are separated by a hyper-plane, the positive vectors are marked by white circle while the negative marked by black circle. In the model, the SVM classifier seeks the optimum hyper-plane which possess the biggest separation margin.

In most of realistic case we find the data points are not linearly separable in the input space, thus the kernel function works perfect for this case. To avoid the possibility to be over-fitting, we can use the kernel function to transform the input space to a higher dimensional space. The kernel function necessarily satisfies the Mercer's condition.

The kernel function cast a significant role of implicitly mapping the input space to a higher dimensional feature space, in this case, separation can be better considering that the original model would lead to over-fitting learning. There are several kinds of kernel functions, such as the most widely used: polynomial kernel function:

$$K(x_i^T + x_j + 1)^p$$, where p is a positive constant.

Or the Gaussian radial basis function (RBF) kernel given by

$$K(x_i + x_j) = \exp\left(-\gamma \left|\left|x_i + x_j\right|\right|^2\right)$$

Sometimes parameterized using:

$$\gamma = 1/2\sigma^2$$

where $\sigma > 0$ is a constant that defines the kernel width. Moreover, there are some other kernel function which can be applied into different cases, such as hyperbolic tangent function. All of these kernel functions must satisfy the Mercer's condition which mentioned above.

Under the usage of kernel function, the discriminant function in a SVM classifier is:

$$f(x) = \sum_{i=1}^{N} \alpha_i y_i K(x_i, x) + b$$

Where $K(x_i, x)$ is the kernel function, $x_i$ are the support vectors determined from the training data, $y_i$ is the class indicator (e.g. +1 and -1 for a two class problem) associated with each $x_i$, N is the number of supporting vectors determined during training process and b is a scalar representing the perpendicular distance of the hyper-plane from origin.

Support vectors are elements of the training set that lie either exactly on or inside the decision boundaries of the SVM classifier. In fact, they consist of those training examples that are most difficult to classify. The SVM classifier uses these support vectors to predict the label (positive or negative) of samples.

*Method #2: Random Forests*

Random forests are an ensemble learning method for classification and regression. Random forests are consists of many single classification trees [11]. Every single tree would give a classification, and the final result depends on the mode of classification results of these decision trees.

Breiman [12] presented the procedure should comply with following algorithms: Assuming that the number of training set is *N*, and we sample *n* cases of *N* in bootstrap method, which means sample with replacement. The number of sample training set is nearly 2/3 to the number of original training set; Constructing classification and regression tree (CART) for every bootstrap training set. These trees are unpruned. Assuming that there are *M* features for every input vector, and we choose *m* best features out of *M* for split use. *m* is the only adjustable parameter of our experiment; As every classification gives a result, then we aggregate it and the mode of classification results is the final classification;

One benefit of bootstrap sampling method is that there are 1/3 of original training set left unselected, and these samples can be applied into prediction. Random forests gives an estimate error of training, which called OOB (out of bag) error. Breiman [12] mentioned that experiment indicates OOB error is unbiased estimate as cross-validation error.

## IV. EXPERIMENTS

*Exp #1: Support Vector Machine Training*

We first use the feature extraction mentioned above to define the different data vector and then construct an input space. These vectors can be labelled with positive or negative, which represent the proteotypic and unobserved peptides. In SVM classification, these vectors are separated by two parts, the first part is used for training and the other part is used for prediction to check the performance of the training model.

For each peptide in the training set, the vectors possess 1088 features as representation. However, the range of features are not comparable. Considering the sample we presented in the article above: the amino acid composition of peptides varies from 0 to 4 while the melting point can up to several hundred centigrade. Due to these disparate ranges, the features corresponding to degree may dominate in the classification and nullify the effects of other features.

To avoid this problem, each feature vector is normalized to -1 to 1. After calculating the features, we generated a matrix $X$. Suppose $X = (x_1, x_2, \ldots, x_k)$, where K is the number of features. $X_k = (x_{1k}, x_{2k}, \ldots, x_{nk})' (1 \leq k \leq K)$, where n is the number of peptides in training set. $\max(X_k)$ is the maximum value in $x_k$ and $\min(X_k)$ is the minimum value in $x_k$. The normalization method is shown as below:

$$x_{(i,k)} = -1 + 2 \frac{x_{(i,k)} - \min(x_k)}{\max(x_k) - \min(x_k)}$$

After normalization, we cast the classification, and we used the prediction accuracy to indicate the performance of the trained model. The accuracy for four platform are as follows:

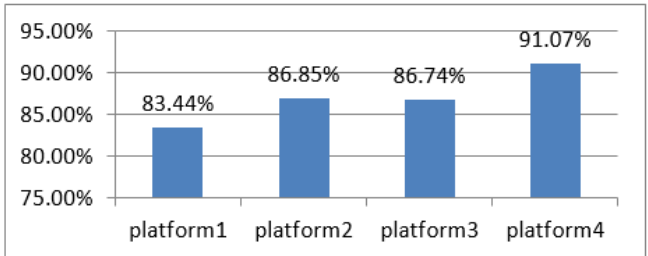

Fig. 3 The accuracy under the SVM classification

The accuracy results shows that the training model has an impressive performance on peptide identification, and some addition work can improve it to an even better results: parameter selection. In our experiments, we cast cross validation accuracy as the index to evaluate which parameter set can lead to the best classification, under the grid search of cost and gamma, we conclude that the cross validation accuracy for four platforms varies from 85%--90%.

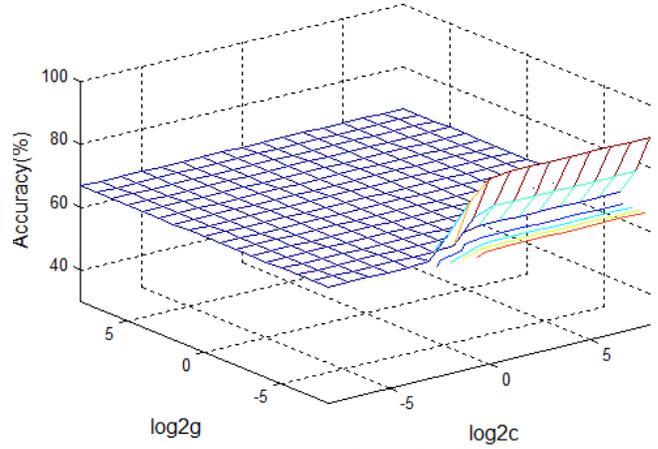

Fig. 4 The parameter selection for platform 4.

The Fig.4 shows the parameter selection for platform 4, and g represents gamma for RBF kernel function and c represents cost function of SVM classification. In this optimum parameters selection, the best c=4 and g=0.0039, the best accuracy is 89%. In this case, the best model has the best robustness.

In conclusion, SVM classifier can lead to a high accuracy of prediction nearly 90%, and the process of parameter selection can improve its performance.

Exp #2: Random Forests Classifier

In this experiment, we employed the random forest method to train a classifier, as it is mentioned above, 100 trees were trained to vote for prediction. And the mtry was set to be 32(the square root of number of 1088 features). The results are listed below:

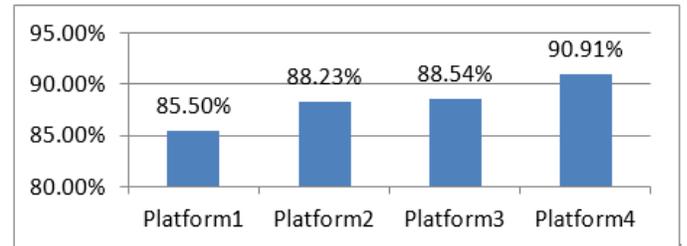

Fig.5 The accuracy under the Random Forest classification

As is mentioned above, the accuracy of Random Forest is the cross validation accuracy, and we can find the accuracy of Random Forest is better than the SVM prediction accuracy, which means that Random Forest is the more suitable classifier for this condition.

Moreover, Random Forest classifier can also provide the function to select features, which is very beneficial for some classification. However, in this case, less features may lead to decrease of accuracy, then we decide to use all of 1088 dimensions. And it is still very convenient considering that computer is completely able to solve the algorithm in few hours.

IV. SUMMARY AND FUTURE WORK

This paper aims at predicting estimate protein abundance by counting the number of observed tryptic peptides. However, the number of observed peptides are usually influenced by their physicochemical property, makes the correction necessary for accurate prediction. In our experiments, we used data from four platform and classify the peptides, then we can obtain a new database to predict new peptide's probability to be detected. In the future, we would try to apply more kinds of algorithms and classifiers to find which one would match this problem most perfectly. Moreover, feature selection is necessary if the accuracy rate can be stable.

V. ACKNOWLEDGEMENT

This work was supported in part by the Nature Science Foundation of China, under Contracts 60903065, 61039003 and 61272052, in part by the Ph.D Programs Foundation of Ministry of Education of China, under Grant 20091102120001, in part by the Fundamental Research Funds for the Central Universities, and by the Program for New Century Excellent Talents in University of Ministry of Education of China. This work is supported by the National High-tech R&D Program projects under Grant No.2011AA010601.